\documentclass[aps,prd,preprint,groupedaddress]{revtex4}

\begin{document}

\title{Comment on ``Singularity-free Cosmological Solutions with 
Non-rotating Perfect Fluids''}

\author{L. Fern\'andez-Jambrina}
\email[]{lfernandez@etsin.upm.es}
\homepage[]{http://debin.etsin.upm.es/lfj.htm}
\affiliation{E.T.S.I. Navales\\ Universidad Polit\'ecnica de Madrid \\ 
Arco de la Victoria s/n \\ E-28040 Madrid, Spain}

\date{\today}

\begin{abstract}
A theorem stated by Raychaudhuri which claims that 
the only physical non-singular cosmological models are comprised in the 
Ruiz-Senovilla family is shown to be incorrect. An explicit 
counterexample is provided and the failure of the argument leading to 
the theorem is explicitly pointed out. \end{abstract}

\maketitle

Since the publication of the first non-singular cosmological model 
with a realistic equation of state \cite{seno}, much effort has been 
devoted to either produce new regular models or to prove that they 
were a set of measure zero in some sense. Raychaudhuri \cite {ray} attempts 
to settle the issue by proving the following theorem:

\noindent \textbf{Theorem:} The only solutions to Einstein equations 
that fulfil the following conditions,

\begin{enumerate}
    \item  Non-singularity: The curvature and physical scalars are 
    regular in the whole spacetime and do not blow up at infinity.

    \item  Perfect fluid: The matter content of the spacetime is a 
    perfect fluid. Therefore the energy-momentum is 
    $T=(p+\rho)u\otimes u+p\,g$, where $u$ is the velocity of the 
    fluid, $p$ is the pressure, $\rho$ is the density and $g$ is the 
    metric.
    
    \item  Non-rotating: The vorticity of the cosmological fluid is 
    zero.

    \item  Cosmological: There is fluid throughout the space which 
    fulfils the energy conditions $0<p\le \rho$. Discontinuities are excluded.

    \item  Changes of $p$ and $\rho$ are in the same direction for 
    any change of the coordinates,
\end{enumerate} 
are those included in the Ruiz-Senovilla family \cite{ruiz}.

A counter example for this claim is supported by the model \cite{leo},
\begin{equation} 
\textrm{d}s^2=\textrm{e}^K(-\textrm{d}t^2+\textrm{d}r^2)+\textrm{e}^{-U}\textrm{d}z^2\,
+\textrm{e}^U\,r^2\,\textrm{d}\phi^2,\end{equation}
which corresponds to a cylindrical cosmological model with matter 
content due to a stiff perfect fluid, $\rho=p$,
\begin{equation} \rho(r,t)=\alpha\,{\rm e}^{-K(t,r)},\end{equation}
\begin{equation} K(t,r)=\frac{1}{2}\,\beta^{2}\,{r}^{4 } + ( \alpha+
    \beta)\,{r}^{2} +2\,{ \beta}\,{t}^{2}
 + 4\,{ \beta}^{2}\,{t}^{2}\,{r}^{2},\end{equation}
 \begin{equation}  U(t,r)={ \beta}\,(\,{r}^{2} + 2\,{t}^{2}\,),
\end{equation} where $\alpha$, $\beta $ are positive constants. The 
coordinates are comoving since the velocity of the fluid is just, 
\begin{equation} {u}= {\rm e}^{ -\frac{1}{2}\,K}\partial_t.
 \end{equation}

The fluid invariants, 
\begin{equation}
{ \Theta}(r,t)=2\,\beta\,t\,(\,1 + 2\,{ \beta}\,{r}^{2}\,)\,{\rm 
e}^{-\frac{1}{2}\,K(t,r)}, \qquad \omega\equiv 0,
\end{equation}\begin{equation}
a^2(r,t)=r^2 \left({ \beta}^{2}\,{r
}^{2} + { \alpha} + { \beta} + 4\,{ \beta}^{2}\,{t}^{2}
 \right)^2\,{\rm e}^{-K(t,r)},
\end{equation}\begin{equation}
\nabla a(r,t)=3\beta^2 r^2\,\textrm{e}^{-K(r,t)},
\end{equation}
\begin{equation}\sigma^2(t,r)=\frac{32}{3}\beta^2\,t^2(1+\beta\,r^2+\beta^2\,r^4)
    \textrm{e}^{-K(t,r)},
\end{equation}are all regular and vanish at spatial and time 
infinity. 

The same happens with the curvature invariants \cite{leo}, 
which are also products of polynomials and decreasing exponentials.

It is obvious that this simple model fulfils Raychaudhuri's 
requirements: It is non-singular and non-rotating, the pressure and 
the density are positive at every point of the spacetime and they are 
related by a state of equation. 

However, it cannot belong to the Ruiz-Senovilla family, since it is 
non-separable in comoving coordinates.

What is wrong then in Raychaudhuri's theorem? The most obvious 
failure in the reasoning leading to his claim lies in the 
decomposition of the pressure that appears in section IV,
\begin{equation}
    p(r,t)=\sum_{i\in I}R_{i}(r)T_i(t)Q_{i}(r,t),
\end{equation} where $R_{i}$, $T_{i}$ vanish at infinity and have null 
derivatives at respectively $r=0$, $t=0$, where the maximum of the 
pressure is reached. 

The author claims that the partial derivatives of the pressure do not 
have any additional zero different from those of $R_{i}$ and $T_{i}$ 
and therefore $Q_{i}$ cannot have any zero. This is obviously false 
and can be illustrated by our model,
\begin{equation}
    p(r,t)=R(r)T(t)Q(r,t),
\end{equation}\begin{equation}
    R(r)=\alpha \textrm{e}^{-\{ ( \alpha+
    \beta)\,{r}^{2}+\beta^{2}\,{r}^{4 }/2 \}},
\end{equation}\begin{equation}
    T(t)= \textrm{e}^{- 2\beta\,{t}^{2}},
\end{equation}\begin{equation}
    Q(r,t)= \textrm{e}^{-4\,{ \beta}^{2}\,{t}^{2}\,{r}^{2}},
\end{equation} which has a $Q$ with derivatives that vanish \emph{also} 
at the zeros of the derivatives of $R$ and $T$, respectively $r=0$, 
$t=0$. 

The author further concludes that $Q$ must be constant, since it is 
bounded everywhere. This would be the case just if analyticity were 
imposed, but this would be  a too restrictive requirement. Our model 
shows again that this claim can be circumvented.

Summarizing, the author's claim clearly leaves aside models that are 
non-separable in comoving coordinates and therefore the issue of the 
existence of further non-singular cosmological models is still open 
within this realm.

In my opinion, separability is too restrictive, since 
$K$ is obtained as a quadrature after the equations for $U$ have been solved 
\cite{manolo}. Even if we impose separability to $U$, integration of 
$K$ will generally lead to cross-terms in $t$ and $r$, breaking the 
separability requirement, as it happens in our model. It is then not surprising that non-separable 
regular models come out if they appear too under the restriction of 
separability.

\begin{acknowledgments}
The present work has been supported by Direcci\'on General de
Ense\~nanza Superior Project PB98-0772. L.F.J. wishes to thank
 F.J. Chinea, L.M. Gonz\'alez-Romero, F. Navarro-L\'erida and  M.J. Pareja 
for valuable discussions. 
\end{acknowledgments}

\end{document}